# Jimbo: A Collaborative IDE with Live Preview

Soroush Ghorashi
School of EECS
Oregon State University
Corvallis, OR, USA 97331
ghorashs@oregonstate.edu

Carlos Jensen
School of EECS
Oregon State University
Corvallis, OR, USA 97331
cjensen@eecs.orst.edu

## ABSTRACT
Team collaboration plays a key role in the success of any multi-user activity. Software engineering is a highly collaborative activity, where multiple developers and designers work together to solve a common problem. Meaningful and effective designer-developer collaboration improves the user experience, which can improve the chances of success for the project. Learning to program is another activity that can be implemented in a more collaborative way, students can learn in an active style by working with others. The growth of online classes, from small structured seminars to massive open online courses (MOOCs), and the isolation and impoverished learning experience some students report in these, points to an urgent need for tools that support remote pair programming in a distributed educational setting.

In this paper, we describe Jimbo, a collaborative integrated development environment (IDE) that we believe is beneficial and effective in both aforementioned activities. Jimbo integrates many features that support better collaboration and communication between designers and developers, to bridge communication gaps and develop mutual understanding. These novel features can improve today's CS education by bringing students closer to each other and their instructors as well as training them to collaborate which is consistent with current practices in software engineering.

## ACM Classification Keywords
D.2.6 **[Programming Environments]:** Integrated Environments;
D.2.2 **[Software Engineering]:** Design Tools and Techniques—*Distributed/Internet based software engineering tools and techniques;*

## 1. INTRODUCTION
The development of software systems is a collaborative process, where team members work together to solve a problem by producing quality code. The designer-developer relationship at the heart of many of these collaborations is the force that moves a software project toward success. Unfortunately, in current software development practices, designers have no direct engagement with developers in the development process, although the products performance depends on both. If we want to improve this relationship, and encourage better software products, we need to build development tools that improve the collaboration and work-flow for designers and developers.

One of the most popular and effective collaboration methods used in CS education is pair programming, which has been shown to be a very beneficial technique for teaching and engaging students with programming and new computing topics. The need for tools that support remote pair programming is becoming pressing with the growing popularity of massive open online courses (MOOC). While employing pair programming in a collocated classroom setting is relatively straightforward, there is a dearth of good options for distributed classroom settings. As students struggle to master concepts and build confidence in their skills, a tight code-artifact feedback loop/mechanism that allows students to verify that a change had the intended result is important.

We have built an IDE, called *Jimbo*, to facilitate the involvement of designers in the development process as first-class citizens. Our focus is on the development of web applications, which often require extensive interactions between designers and developers. In this paper, we briefly describe our tool, its main design goals, and then describe two major collaboration scenarios in which Jimbo can be effective and useful.

## 2. JIMBO
Jimbo is an IDE that enables users to more easily collaborate around a common project. We have tried to make the user interface easy to learn and memorable, but have also considered external consistency with other popular IDEs.

Jimbo is a web-based tool, which means that users only need a standard web browser and the setup time is zero. Sarma provides a comprehensive classification of collaborative tools for software development [1]; Jimbo could be considered a seamless tool at the top level as it provides many novel features to automate the development workflow and minimize user effort. In the following sections we describe these features briefly.

### 2.1 Synchronous and Asynchronous Collaboration.
Currently, distributed collaborative software development largely revolves around working in parallel on separate copies of the code, and integrating the resulting efforts using a source code version control system. Though this is an effective strategy, there is a lack of real collaboration, as developers are largely working independently, only coordinating their efforts when synching their code. The number of defects in code tends to rise as the amount of parallel work increases [2] and developers sometimes avoid this kind of development to avoid having to resolve conflicts [3].

The most important feature of Jimbo is that it allows developers to collaborate on code in real-time without worrying about conflicts. Jimbo deploys an operational transformation (OT) algorithm to achieve this. OT is a technique that provides eventual consistency between multiple users working on the same code without retries, errors, or data being overwritten.

### 2.2 Communications
The integration of communication features into the IDE helps collaborators discuss and resolve issues without losing focus on

the code. We recommend the following different but equally effective means of communication:

1. **Audio chat.** This provides virtual presence and makes it easier to coordinate and collaborate. This type of communication allows for quicker resolution.
2. **Text chat**. A text-based chat or instant messaging is a complimentary system for audio chat and is a must, as coders share code snippets and links to resources.
3. **Inline discussions.** Inline discussions are text-based semi-synchronous way of communication, in which users can associate a short discussion with an artifact such as a portion of code or a design document.

## 2.3 Live Preview

Live programming is a technique where programmers re-execute a program continuously while editing [4]. Live preview is a variation of this technique that refreshes the output immediately upon a change to the code, and it best fits UI-heavy application development such as websites. This technique provides an immediate connection between the code and the output for developers so they can see the effects of changes to their code. This feature leads to fewer iteration of the code, which means faster coding.

Currently web programming requires developers to write their code in an editor and then open it in a standard browser to see the results. Any changes they make to the code requires them to save the file, go to their browser and refresh the page to check the effect of changes made. When this is done collaboratively, the files edited have to be pushed to a server, and potential conflicts have to be managed, all of which is slow and effort intensive. Jimbo provides a real-time preview of the code being developed, which streamlines the development process by removing redundant refreshes on the browser.

## 2.4 User Awareness

Another fundamental requirement to supporting collaboration is awareness. Dourish and Belloti define awareness as "an understanding of the activities of others that provides a context for your own activity" [5]. The main purpose of an awareness system is to provide information about development activities to help coordinate tasks. In Jimbo, we follow the "continuous coordination" model introduced by van der Hoek et al. [6]. The primary responsibility of such a system is to notify developers of events relevant to them, such as code changes, comments to discussion threads, user presence, etc.

Jimbo has a powerful channel based notification system [7] using push notifications. These notifications are persistent and stored on the server for future retrieval. To prevent cognitive overload, developers can request to receive notification about a specific portion of the code, a feature we call "code watch". Once someone puts a watch on a portion of code, Jimbo pushes notifications regarding any changes to that section of code. This feature allows developers to keep track of sections of code their work depends on, or code that they have some owner- ship over.

## 3. COLLABORATION SCENARIOS

We have identified two main populations that can benefit from a tool like what we have described. We have therefore created two versions of Jimbo: One educational (Figure 1) and a professional (Figure 2). First, we discuss how Jimbo can help students and instructors, then we explain how designer-developer collaboration can be improved using Jimbo.

## 3.1 Educational Settings

One of the most popular models of collaboration is pair programming. The impact of pair programming in CS education has been confirmed by numerous studies in both lab and classroom settings [8, 9, 10, 11]. Pair programming requires two programmers to work together on the same computer, but the trend toward geographically distributed teams make long-distance adaptations necessary. A modified model in which collaborators are not collocated is called remote pair programming (RPP), and it has been shown that RPP provides the same benefits as collocated pair programming [12]. While it is important to facilitate both synchronous and asynchronous collaboration – providing the most flexibility for students, synchronous collaboration is perhaps the most important as it makes it possible for students to engage in RPP at any time and from any location without worrying about syntactic consistency.

The integration of communication features into the IDE could help students discuss without losing focus on the code. Most communication tools (Skype, IRC, email, etc.) are not directly integrated with the IDE, which can lead to a disconnect between code and discussion, or simply wasted effort by frequent context switching. A text-based chat system is a must, as students share code snippets and links to resources. An audio chat system allows students to talk freely similar to effective face-to-face meetings.

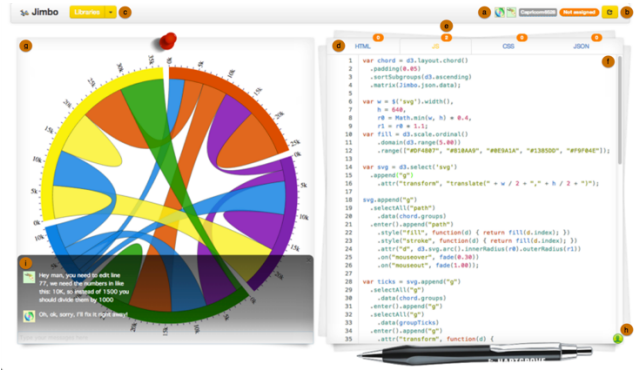

**Figure 1. Jimbo educational edition.**

Another popular practice is "live coding," where an instructor writes code in front of students, exposing their thought process. While this is more effective than just showing the final solution to students, it is a passive learning technique. Gaspar and Langevin [13] successfully used a student-led version of this approach to engage students in active learning and expose students' thought process to the instructor for more in-depth feedback. To better support student-led live coding and engage all students during a class session, having a tight code-artifact feedback loop that allows students to verify that a change had the intended result is important. One advantage of combining a live preview component with pair programming is that live preview supports a distributed and thus scalable way of engaging in live coding. This would allow all students to benefit from this practice during short class periods, whereas before only a handful of them would get the chance to practice this technique.

Figure 1 shows the main view of the Jimbo educational edition. To simplify the programming process for students who are new to the web development topics, there are exactly one editor for each of the HTML, CSS and JavaScript codes and they are located on the right side of the view. Live preview, if enabled, can be found on the left side of the view. Text chat panel is located on the bottom left and students can join an audio call by simply pressing a button which is positioned on the top right.

## 3.2 Professional Settings

Software teams usually have more than one developer and designer collaborating. Instead of isolating designers from the development, we suggest including them in the implementation process, working with the developers while these work with the code. Many designers have basic development skills, most notably HTML, CSS and JavaScript for prototyping. Having synchronous coding integrated into IDE tools enables designers to make modifications directly, or work with the developers and direct their development efforts without going through a long set of steps to deliver their ideas to the engineering team effectively. This can facilitate a more design oriented process in which designers create the skeleton of the product and then developers fill in the gaps to bring the design to the life. This way, designers guard their designs themselves and developers are forced into the habit of getting more immediate and meaningful feedback.

The integration of communication features into the IDE could help developers and designers discuss and resolve the issues they may face without losing focus on the code. Audio chat allows for quicker resolution in case of minor design misunderstanding between developers and designers that if stay unresolved, it sometimes will lead to major defects in the product and delay in the process and in worst case failure of the project. Unlike transitory audio discussions, inline discussions are tied to people and artifacts and tend to be permanent, having their own context that will not be lost over time.

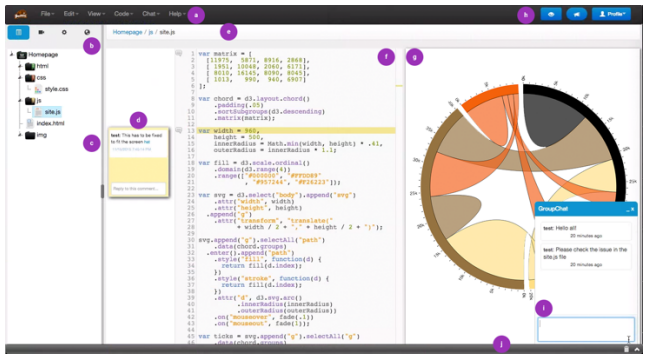

**Figure 2. Jimbo professional edition.**

In an environment where designers and developers are collaborating with each other, live preview is a powerful tool. Using live preview, designers can instantly see what changes developers are making and provide feedback and direction, warning the coding team if they are deviating from the design vision. It also streamlines the flow of communication within the team, and makes it easier for developers to ask designers for input or help when changes have to be made.

Awareness is required to coordinate teams, but can be distracting if it interrupts or requires too much attention from members. It is not an easy issue to address in a collaborative environment, as we juggle the need for asynchronous editing for some developers, and the need for real-time preview of the resulting code for others. A good awareness system in such environment will provide useful answers to the questions that a user may have, for example: Who made what change in the code? Who should I contact if I have a question about this part of design/code? Who is currently available? What are they doing now?

Figure 2 shows the main view of the Jimbo professional edition. The UI in this version is similar to popular IDEs such as Eclipse, with the project file tree on the left, and code editor expanding to the right. The text chat panel can be found on the bottom right of the screen, and inline discussions are located on the right edge of the code editor. If a user wishes to have their live preview enabled, they can find its panel on the right side of the view. Finally, notification center, audio conference call and user profile options are located on the top right of the screen.

## 4. RELATED WORK

Researches have studied collaborative coding and built many tools following the WYSIWIS (what you see is what I see) metaphor that supports collaboration, including: GROVE [14], ShrEdit [15], DistEdit [16] and Flesce [17]. One of the key challenges for these tools is making sure that the code is always error-free, as any collaborator may run and test the code at any point in time, regardless of what the others are doing. Collabode [18] has been developed to address this issue by only sharing modifications that result in an error-free code, acting as an automatic safety buffer.

A number of tools have been developed to provide real-time awareness of code changes to facilitate coordination and emerging conflicts in collaborative environments. FASTDash [19], ProjectWatcher [20], Palantír [21] and Syde [22] are all examples of this kind of tools. Crystal [23] proactively watches the code and precisely identifies and reports conflicts. There are also growing number of plug-in services being developed for existing IDE's, trying to add more awareness and collaboration features to familiar tools such as Eclipse JAZZ project [24].

Some recent live programming systems include Superglue, Flogo II, Lively Wiki and Brackets [25, 26, 27, 28]. In a real world example, Khan academy recently successfully deployed a basic version of a live programming environment for an online programming course for students with no experience [29]. Researchers are trying to enhance live programming environments by focusing on debugging [30].

## 5. CONCLUSION AND FUTURE WORK

In this paper, we introduced Jimbo, a web based collaborative IDE with live preview built mainly for web development. Jimbo integrates a number of techniques and features, including synchronous coding, user awareness, live preview and various communication channels that previously existed in separate tools. We briefly discussed two main user scenarios – the educational and professional – that we think Jimbo can be most beneficial in.

We have developed multiple editions of Jimbo to support collaboration in educational settings where students thrive on collaboration in problem solving, and also professional settings where people in different roles try to solve a common problem collaboratively. Currently we are conducting multiple user studies following these two settings to investigate whether our tool Jimbo; truly helps users to solve real-life problems effectively, and we look forward to reporting the results of these studies.

## 6. REFERENCES

[1] Sarma, A. A survey of collaborative tools in software development. Technical Report UCI-ISR-05-3, University of California, Irvine, Institute for Software Research, 2005.

[2] Perry, D. Siy, H. and Votta, L. Parallel changes in large-scale software development: an observational case study. ACM TOSEM, 10:308–337, July 2001.